%% file: main.tex
\documentclass{article}

\usepackage{fullpage}
\usepackage{authblk}
\usepackage[round]{natbib} 
\usepackage{color}

\usepackage[utf8]{inputenc} 
\usepackage[T1]{fontenc}    
\usepackage{hyperref}       
\usepackage{url}            
\usepackage{booktabs}       
\usepackage{amsfonts}       
\usepackage{nicefrac}       
\usepackage{microtype}      
\usepackage{amsmath}
\usepackage{color}
\usepackage{amssymb}        
\usepackage{caption} 
\usepackage{wrapfig}
\usepackage[pdftex]{graphicx}
\usepackage{subcaption}
\usepackage{amsmath}
\usepackage[title]{appendix}
\usepackage{mathtools}
\usepackage{pgf, tikz}
\usetikzlibrary{arrows, automata}
\DeclarePairedDelimiterX{\infdivx}[2]{(}{)}{%
  #1\;\delimsize\|\;#2%
}

\renewcommand{\vec}[1]{\mathbf{#1}}
\newcommand{\distas}[1]{\mathbin{\overset{#1}{\kern\z@\sim}}}%

\title{Factor Representation and Decision Making in Stock Markets Using Deep Reinforcement Learning}

\author{
  Zhaolu Dong ~~~ Shan Huang ~~~ Simiao Ma ~~~ Yining Qian \\
  Georgia Institute of Technology \\
  correpondence to:  shuang418@gatech.edu
  
}

\begin{document}
\maketitle

\begin{abstract}
Deep Reinforcement learning is a branch of unsupervised learning in which an agent learns to act based on 
environment state in order to maximize its total reward. Deep reinforcement learning provides good opportunity to model the complexity of portfolio choice in high-dimensional and data-driven environment by leveraging the powerful representation of deep neural networks.
In this paper, we build a portfolio management system using direct deep reinforcement learning to make optimal portfolio choice periodically among S\&P500 underlying stocks by learning a good factor representation (as input). The result shows that an effective learning of market conditions and optimal portfolio allocations can significantly outperform the average market.  
\end{abstract}

\input{tex/method}

\clearpage
\bibliography{mybib}
\bibliographystyle{mybib}

\end{document}

%% file: tex/method.tex
\section{Literature Review}
Both deep learning (DL) and reinforcement learning (RL) are biological inspired frameworks and theoretically rooted in the neuroscientific field for behavior control. In practice, they get more and more attention in finance applications. One actively researched area is about applying such models for market forecasting. The literature review regarding this field is recently done in \cite{article1}. Another area is to solve dynamic financial decision-making problems or train relative trading systems via these models. 
For instance, Jiang et al. \cite{article9} use the model-free Deep Deterministic Policy Gradient (DDPG) \cite{article12} to dynamically optimize cryptocurrency portfolios. Similarly, Liang et al. \cite{article10} optimize stock portfolios by using the DDPG as well as the Proximal Policy Optimization (PPO) \cite{article11}.
Fister et al. \cite{article3} train trading system by deep LSTM while Moody and Saffell \cite{article2} by recurrent reinforcement learning.

Deng et al.\cite{article4} implement DL to automatically sense the dynamic market condition for informative feature learning and RL to make trading decisions. To the best of our knowledge, \cite{article4} is the first paper to learn (future) market conditions and trading decisions at the same time, i.e., the two areas as mentioned above. However, \cite{article4} only handles one share of the asset. Building upon this, we will extend the DL and RL framework learned in a complex neural network (NN) to a portfolio setting, extracting features from multiple assets and then making optimal trading decisions.\

In reinforcement learning, an agent must be able to sense the environment, make decisions, and achieve goals \cite{article5}. A typical reinforcement learning problem can be generally categorized into value-function-based method and actor-critic-method \cite{article6}. A value-function-based method, such as Q-learning \cite{article7}, can solve  the  optimization  problems  defined  in  a  discrete  space by estimating a value function like future discounted return. The applications of Q-learning in the portfolio optimization problem can be found in \cite{article13} \cite{article14} \cite{article15}. 
However, this method is not ideal to solve trading problem because the trading environment is more complex than a discrete space. Unlike the value-function-based method, in an actor-critic method, the “critic” learns a value function and then uses the result to update the parameters of the “actor.” The actor-critic based methods such as DDPG and PPO for portfolio optimization problem can be found in \cite{article9} and \cite{article16}. Another example of actor-critic-method is direct reinforcement learning that learns from the continuous sensory data. Moody and Saffell \cite{article2} found that RRL direct reinforcement framework eliminates building a forecasting model, avoids Bellman's curse of dimensionality, and provides better results than Q-learning \cite{article8}.

\section{Problem Definition}
Portfolio management is the action of continuous reallocation of a capital into a number
of financial assets. For an automatic portfolio management system, these investment decisions and actions
are made periodically based on investment information states. This section provides a mathematical setting of the portfolio
management problem.

\subsection{Trading Period}
In this project, we consider daily stock tradings so that the trading agent reallocates his asset allocations once per trading day. For each trading day, we have the 
basic market information, namely the opening, highest, lowest and closing prices, as well as the trading volumes. Extra integrated market information includes manually constructed factors using historical 
market and news information. It is assumed in the back-test experiments that in each trading day stocks can be bought or sold at the closing price. We only consider long positions since short selling is risky.

\subsection{Data}
The agent manage portfolios among the S\&P500 underlying stocks. The historical daily price information (up to today) can be obtained using module ``tidyquant'' in R. To use as many information as possible, we also include several indicators for technical analysis such as EMA, RSI, MACD and so on. Further, we also include macro economic features and sentiment of financial news.

\subsection{Mathematical Formalism}
We construct portfolios of one risk-free asset (bond) and $m$ stocks. At trading day $t$, the risk-free asset earns risk-free rate of return $r^f_t$, and the $i$th stock has price $P_{i,t}$ for $i=1,2,...,m$. The market return of $i$th stock for the period ending at time $t$ is defined as $\left(\frac{P_{i,t}}{P_{i,t-1}}-1\right)$. Defining portfolio weights of the $i$th stock at time $t$ as $\omega_{i,t}$, the agent that only take long positions have portfolio weights that satisfy
\begin{equation}
    \sum_{i=0}^m \omega_{i,t}=1,\text{ and }  \omega_{i,t} \geq 0, \ \ \forall t\geq 0, i=0,1,2,...,m,
\end{equation}
where $\omega_{0,t}$ represents the weight allocated in bond. 
By the equation above, it is a natural idea to treat the portfolio weights as the outcome of a neural network in which the output activation function is a softmax function. 

When multiple assets are considered, the effective portfolio weights change per trading day due to price movements. Thus, maintaining constant or desired portfolio weights requires that adjustments in positions be made at each trading day. If there is no transaction costs, The wealth after $T$ periods for a portfolio management system is
\begin{equation}
    W_T = W_0 \prod_{t=1}^T (1+R_t) = W_0\prod_{t=1}^T \left(\sum_{i=0}^m \omega_{i,t} \frac{P_{i,t}}{P_{i,t-1}} \right),
\end{equation}
where $R_t$ is the portfolio return at time $t$ which depends on price change and previous allocations, $P_{0,t}$ is the bond price and by the risk-free property we have $P_{0,t}/P_{0,t-1}=1+r_f$. 

In real world, transaction costs are not negligible for frequent tradings. We assume a proportional transaction cost $\delta$ is charged whenever rebalancing positions. Then the wealth after $T$ periods becomes
\begin{equation*}\begin{aligned}
    W_T =& \ W_0\prod_{t=1}^T \left(\sum_{i=0}^m \omega_{i,t} \frac{P_{i,t}}{P_{i,t-1}} \right) \left( 1- \delta \sum_{i=1}^m \big | \omega_{i,t} - \tilde\omega_{i,t} \big | \right), \\
   \tilde\omega_{i,t} =& \ \frac{\omega_{i,t-1}P_{i,t}/P_{i,t-1}}{\sum_{j=0}^m \omega_{j,t-1} P_{j,t}/P_{j,t-1} },
\end{aligned}\end{equation*}
where $\tilde\omega_{i,t}$ is the effective portfolio weight of stock $i$ before readjusting at time $t$.

At the beginning of each time $t$, base on the observations $\{(P_{i,s},\omega_{i,s}): s\leq t, i=1,2,...,m\}$ the agent tries to find out the optimal portfolio weights $\omega_{i,t}$ that can maximize the final portfolio wealth. Thus, the objective is
\begin{equation}
    \max_{\{\omega_{i,t}: \ t=1,2,...,T; i=1,2,...,m\}} \mathbb{E} \left[ U_T(R_1,R_2,...,R_T) \right],
\end{equation}
where $U_T$ is the utility. The most common one is to choose $U_T(R_1,R_2,...,R_T)$ as log utility of final  wealth so that equivalently the objective function is to maximize the cumulative portfolio returns $\sum_{t=1}^T R_t$. Other utility function such as portfolios' sharp ratio can be considered. 

\begin{equation*}
    \max_{\{\omega_{i,t}: \ t=1,2,...,T; i=1,2,...,m\}} \frac{avg(R_1,R_2,...,R_T)}{std(R_1,R_2,...,R_T)},
\end{equation*}
where the standard deviation of returns in the denominator can also be substituted by drawdowns to further control risks.  

\subsection{Factor Representation}
Vintage data input is important for training DRL. Usually the financial data contain a large amount of noise, jump, and movement leading to the highly nonstationary time series. To learn robust feature representations directly from data, we borrow the idea from the development of Natural Language Processing, i.e., to use GRU or LSTM network to learn the integrated feature representation of historical data. The 
Drop out method used based on RNN can help to find the underlying features and reduce the uncertainty of data from \cite{han2017stein}, \cite{han2018stein} and \cite{ han2020stein}. 

\subsection{Direct Reinforcement Learning }
Typical Reinforcement Learning can be generalized into two types as critic-based (learning value functions) and actor-based (learning actions) methods.  Critic models are algorithms that directly estimate value functions and solve problems defined in discrete space using dynamic programming. Among the available algorithms, Q-learning is the most widely used and it is the first method to be adopted in our project. However, just like it is mentioned above, Q learning failed to approximate the complex market dynamics. At the same time, the calculations of value function usually require re-coding of future discounted return. In reality, when we deploy the RL system in online learning\& trading mode, information of future return apparently does not exist for computation and hence, Critic models are not ideal for trading cases.\

For actor based RL models, the agent directly evaluates trading policy (instead of valuing market conditions in discrete states) defined in terms of continuous action space. Without using dynamic programming, the optimization functions simply require a differentiable function and a few latent variables, which saves a considerable amount of calculation power. In addition, the actor based model is capable of handling continuous data. Thus, it is advantageous to handle financial market dynamics, especially in the case of practical applications in online mode. Specifically, Direct Reinforcement Learning is used in our final model.\\

With the well-defined mathematical formalism of utility function, an efficient strategy to learn the trading policy directly is by using Direct Reinforcement Learning. Specifically, we approximate the action at each time period as the weight allocated on each asset within the portfolio.
\begin{equation}
     \vec{w}_{t}=(w^{1}_{t},w^{2}_{t}...w^{n}_{t}),
     w^{i}_{t}>=0
\end{equation}

using nonlinear equation:
\begin{equation}
     \vec{w}_{t}=RDNN[f_{t},\vec{w}_{t-1}]
\end{equation}
where RDNN represents a Recurrent Deep Neural Net manipulation, $f_{t}$ represents feature vector of market information at time t; we added the previous action term $u\vec{w}_{t-1}$ in computation in order to consider the previous action and discourage frequent change in position to avoid potential heavy transaction cost. After RDNN, the output maps the function to final asset allocation in the range of (0,1). Note that, with the decided allocation of asset at time t, the amount of stock we need to trade for asset i will be
\begin{equation}
     \bigtriangleup stock^{i} =\frac{wealth_{t}*w^{i}_{t}-wealth_{t}*w^{i}_{t-1}}{price^{i}_{t}}
\end{equation}
In conclusion, Direct Reinforcement Learning aims to study the set of trading rules within different state (associated with feature input and previous actions) in order to maximize the expected utility function.

\subsection{RNN,DL and RDNN}
 In reality, the price movement of stocks is dependent on its last price and trades. So the analysis of stocks should use its reasoning about previous events to inform the later ones based on \cite{han2016bootstrap}, \cite{han2017high} and  \cite{han2021disentangled}. Such a sequential manner can be facilitated by Recurrent Neural Network as it helps to persist long-short time information. A recurrent neural network can be understood as loops of the same networks each passing a message to its successor. In our project, we took advantage of RNN and constructed a one layer GRU (a type of RNN) for fuzzy feature representation,as it links the previous trading allocation $\vec{w}_{t-1}$ to the input layer in order to discourage frequent trading.This process can be formulated as follows.
\begin{equation}
     \vec{o}_{1}=RELU[<a,f_{t}>+b+u\vec{w}_{t-1}]
\end{equation}
where$\vec{o}_{1}$ represents output from the first layer of GRU, <,>represents the inner product between two input variables, a,b and u represents coefficients for feature regression. 
After linear regression, the piece-wise linear method of
\begin{equation}
     RELU(x)=MAX(0,x)
\end{equation}
is used as activation function to map the function to the output of first layer with value no less than zero.\\

In addition, Deep Learning(DL) is a machine learning algorithm inspired by the structure and function of the brain called artificial neural networks  \cite{lombardo2019deep}, \cite{han2018deep}. With a layer to layer structure to hierarchically transform information, Deep Learning is proven success in learning behaviours and optimizing decision making. In our project, one drawback of a one layer NN is the lack of a "brain" to sense market shift and adjust its trading decision. Therefore, we introduce Deep Learning in our DRL model to study trading behaviour with policy gradient for optimized decision making. \\
Regarding deep representation, the deep neural network has multiple connected layers to hierarchically transform input to output vector. Based on the first layer of GRU network, we further constructed a two-layer DNN to map the output of fuzzy feature representation into final decision of asset allocation. The output of the ith node in layer l-1 $o^{i}_{l-1}$ is now made to be the input in layer l. The transformation between layer l-1 and layer l is 
\begin{equation}
     z^{i}_{l}=<a_{l}^{i},o^{l-1}>+b^{i}_{l}
\end{equation}
with $z^{i}_{l}$ being the intermediate value from linear regression and
\begin{equation}
    o^{i}_{l}=Activation(z^{i}_{l})
\end{equation}
Where the parameter set $<a^{i}_{l}, b^{i}_{l}>$ are the layer-wise latent variables for DNN to learn. Note that the second (which is the first layer of DNN) layer of the overall neural network computes output by using RELU method as the activation function, which is the same as the previous layer of RNN; while the last layer computed the output as final asset allocation decisions by using Sigmoid as the activation function. 
\begin{equation}
    Sigmoid(x)= \frac{e^x}{e^x+1}
\end{equation}
Note that the value of the output using sigmoid function is ranging from 0 to 1, hence obey the rules of no short-selling.

In conclusion, the overall structure of our DRL model consist of a one-layer RNN for feature representation and a two-layer DNN for studying trading actions. Note that the DNN uses extended features which includes previous actions. Thus, we can call the DNN learning as RDNN. According to the research conclusion, a neural network consists of 128-128-64 neuron is the most efficient structure for the case of financial analysis. Hence, it is also adopted in this project.

\subsection{Task-Aware Back Propagation Through Time }
To frequently adjusted to market change, updates of parameters are required with back propagation (gradient decent) when a batch of data is passing. However, there are issues of back propagation with recurrent and deep structures. If we denote $\theta$ as the general parameter in RDNN model,then its gradient can be calculated using chain rule
\begin{equation}
     \frac{\partial U_{t}}{\partial \theta}=\Sigma_{t}\frac{d U_{t}}{d R_{t}} \{\frac{d R_{t}}{d w_{t}} \frac{d w_{t}}{d \theta}+\frac{d R_{t}}{d w_{t-1}} \frac{d w_{t-1}}{d \theta}\}
\end{equation}
\begin{equation}
    \frac{d w_{t}}{d \theta} =  \frac{d w_{t}}{d h^{2}_{t}}\frac{d h^{2}_{t}}{d h^{1}_{t}}(\frac{d h^{1}_{t}}{d \theta }+\frac{d h^{1}_{t}}{d w_{t-1}}\frac{d w_{t-1}}{d \theta})
\end{equation}
where $h^{i}_{t}$ represents the unlinear mapping of function through the ith layer of Neural Network at time t.
Note that as we also include the previous asset allocation $w_{t-1}$ as our input,when deriving $d w_{t}/d \theta$, a subsequent derivation of $d w_{t-\tau}/d \theta$ $\tau =1....T$ is calculated recursively,imposing difficulties on gradient descent. By setting a time stack based on different values of $\tau$ for time-based unfolding, the current system reduces to a minimal recurrent structure and the typical BP method is easily applied to it. In addition, parameters' gradients at each time step are averaged together to form the final gradients.\\
However, the original DNN now gets deeper and with time stacks, leading to issues of gradient vanishing. To solve this, we adopt a method of setting up virtual connections between policy function and variables of each time stack to bring more gradient information.

\subsection{Methods}
In this paper, we use python for coding along with templates including Pandas, Numpy and Pytorch.

First of all, we set up a financial environment with initial capital of \$100,000 in cash and a commission rate of 0.01\% based on amount of trade. Trading frequency is set on daily bases and no short selling is allowed. Data includes information of daily adjusted open, closed, high, low prices and volumes of stocks and ETFs from 2007 to 2019. We select this time range because it covers enough information to demonstrate shifts between bull and bear markets. Then, we calculate technical indicators and vectorize all available information as state for agent to train. Finally, to stabilize and smoothen price time series, all the information is normalized with data in the past 20 days.

In the next step, we build a deep-reinforcement-learning-driven trading agent. We first apply drop-out method, where randomly selected neurons are ignored during training to enhance model robustness. The dropout rate is set to be 20\%. We then initialize the first layer of DRL network, which is a recurrent net with 128 neurons and with RELU as activation function.  After the RNN process, we build two two-layer DNN models with 128 and 64 neurons respectively, with activation function of RELU for the first layer and Sigmoid for the second layer. The learning rate is set to 0.001 and batch size is set to 64.

In order to compare the performance of our system with respect to bench mark index of S\&P500, we select an ETF called SPY, which is the largest ETF to track S\&P500, as our single portfolio for the first strategy. Then we trained a portfolio consist of picked stocks and one risk-free asset. After that, we compare the results of total profit between these two strategies.

\section{Result and Discussion}
Results of the performance of strategy with consistent training \& trading sequences show strong profit making ability by outperforming the strategy of buying and holding (to simulate return from S\&P500) by over \$250,000 dollars in around 10 years.
\begin{center}
  \includegraphics[scale=0.7]{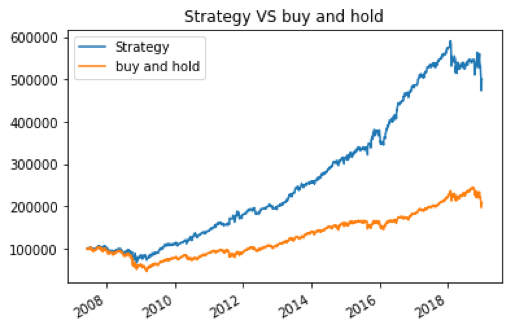}\\
  Figure 1 :Performance comparison between agent’s strategy with the buying and holding strategy in terms of investment return (\$)
\end{center}
At the same time, it is worthwhile pointing out from Figure 1 that, during the period of financial crisis (between late 2008 and early 2009), the value between two sets of portfolio start to diverge, showing that the Deep Reinforcement Learning system has developed senses of risk and corresponding action, thus experiencing less loss. \\
To further investigate the strategy used by DRL under different market environment, we trained a system agent under the period of financial crisis, then stop training it to preserve the strategy and let it trade in the following period.  As shown in Figure 2, portfolio value with this strategy stays in a stable zone during the time when market experience sharp decline in March 2009 and March 2010. However, as market start to rise in late 2010, conservative holdings of shares will lead to value of portfolio being surpassed by holding strategy.
\begin{center}
  \includegraphics[scale=0.9]{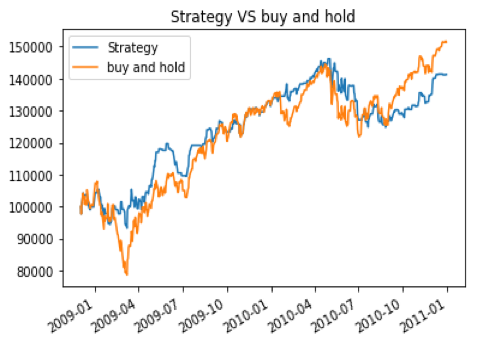}\\
  Figure 2 :Trading performance of risk averse strategy
\end{center}
On the other hand, we guided the agent to develop a risk-seeking strategy by training it during early 2006, which is the rising period. We evaluated this performance during the period of financial crisis. As expected, the aggressive holding of  portfolio leads to an under-performance comparing to benchmark. As indicated by Figure 3, the portfolio value of our strategy stays under buying and holding strategy under periods of fluctuations and market collapse.
\begin{center}
  \includegraphics[scale=0.9]{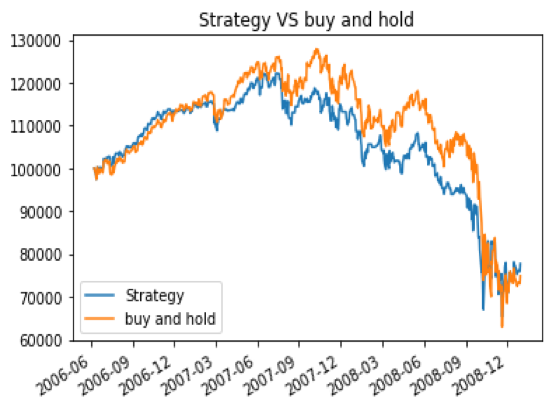}\\
  Figure 3 :Trading performance of risk seeking strategy
\end{center}
Finally,the shift in agents' trading appetite can be future proved in the figure of Gross Leverage, as shown below. During year 2006 to 2007, the gross leverage line stays at around 1 (100\%), meaning that our trading agent have all the wealth allocated on stocks. However, after experiencing the great recession in 2008, our agent's trading behaviour became more conservative, with holding less stock on hand (shown as the troughs between year 2008 to year 2010) when market is volatile.  
\begin{center}
  \includegraphics[scale=0.8]{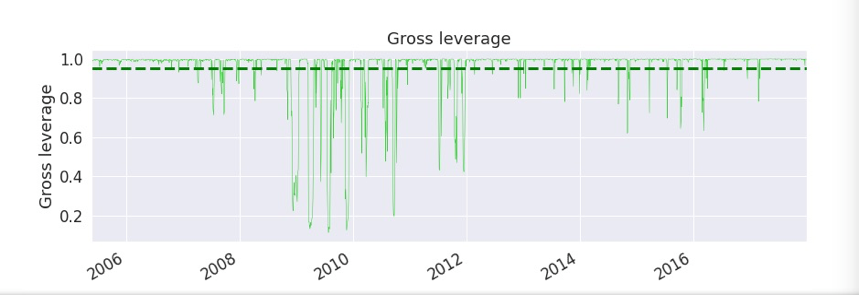}\\
  Figure 4 :Gross Leverage (Proportion of wealth invested in stocks)
\end{center}

\section{Limitations and Next-Step Improvements}

The first limitation is that our model failed to consider market impacts of trades on stocks. As we have mentioned in the beginning, we assume zero market impact assumption. That is, the agent we trained is a small market participant and the capital invested by is so insignificant that is has no influence on the market. Though, price impact (either temporary or permanent) is implemented in classical structural model. It is difficult to implement it in our model since our model is only data-driven and we cannot rewrite the historical data. 

The second limitation associated with our model is that, with the increase of trading cost, the performance of Deep Direct Reinforcement Learning reduced significantly. To address this issue, the next step of our implementation is to add constraints on volume turnovers. For example, we can set a threshold on the turnovers in the basket. Future, we will try different definition of actions actions in our DRL. Currently our control is the portfolio weight. In the presence of large transaction costs, this may not be suitable since the portfolio weights naturally change with price movements. Hence, decreasing the cost of transactions needs the action we learned to be partially predictable on stock movement, which is difficult. To make it earlier, we will try to directly control the number of shares invested in each stocks. Therefore, decreasing the cost of transactions is simply equivalent to decrease the change of stock shares. Similarly, we can add constrains on turnovers. For instance, we can restrict half of the stock tickers in the basket unchanged.

Thirdly, our model only consider limited pre-selected stocks. Currently we only model the optimal portfolio allocations given 20 pre-selected stocks. By fixing the total number of stocks in portfolio, in the future we may try general optimal stock selections directly from the underlying stocks in S\&P500.  That means, the stock tickers in our basket may change over time. Note that increasing the number of stocks will cause exponential growth of calculations. As we add in more stocks of interest, the resulted wealth achieved by the Deep Reinforcement Learning system downgrades significantly due to the costs in high trading frequency and aggressiveness. By controlling the turnovers as mentioned above, we hope the new general portfolio selection model can generate promising results. The further implementations are done in the following subsections.


\subsection{Mask Network for general selection}
We introduce mask for general selection. To select a fixed number of stocks, see e.g., 20 stocks, from a large pool of stocks, we need to introduce a mask vector, in which the value 1 means select and 0 means non-select. This mask vector is generated by a different network separated from actor-NN. As shown in the figures below. The mask network takes the state variables of all stocks in the large pool as input and generate a score vector, then the mask is generated by selecting the top 20 scores. Note that at time 0 and 1, the ticker names of top 20 stocks selected may be totally different (as distinguished by different colors). Then the normalization of the multiplication of actions generated by DRNN and the mask gives the portfolio weights. We can say that the portfolio weights of the 20 selected stocks are non-zero, while the portfolio weights of the remaining stocks are all zero. 

\begin{center}
  \includegraphics[scale=0.65]{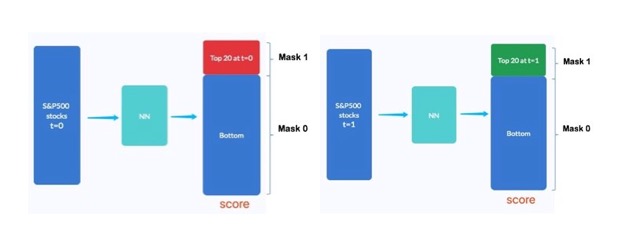}\\
  Figure 5 :Mask network for general stock selection.
\end{center}

\subsection{Turnover Control}
The application of mask can help us make general selections on large pool of stocks, but it contains lots of uncertainty. As we have mentioned. The stock tickers selected at time 0 may be totally different from those selected at time 1. In the case at time 1, our agent needs to first liquidate all stocks invested at time 0. This liquidation usually generates large transaction costs. To solve this issue, we further add constraint on the turnover in our portfolio. We can restrict the turnover of stocks no more than a given threshold.  For example, we can fix half of the stock tickers in our basket unchanged in two consecutive trading day. The tricky part is that we need to guarantee half of the stock tickers in the basket come from the selections in previous trading day. In the shown figure, at time 1, the mask network generates a score vector. We only need to select the top 10 additional stocks since the top 10 stock tickers selected at time 0 must be reserved at time 1. Using the score vector generated at time 1 and the 20 stock tickers selected at time 1, we can update the 10 tickers that will be reversed at time 2. Similar process goes to the whole time period. The comparison result of general selection model with and without turnover control is shown in this figure. The green line is the average market performance using simply buy and holding strategy. The blue line shows the general auto trading result on the pool of S\&P 500 underlying stocks without turnover control, while the orange line shows the result with 50\% turnover control. It is seen that turnover control deed help increase our model performance. 

\begin{center}
  \includegraphics[scale=0.8]{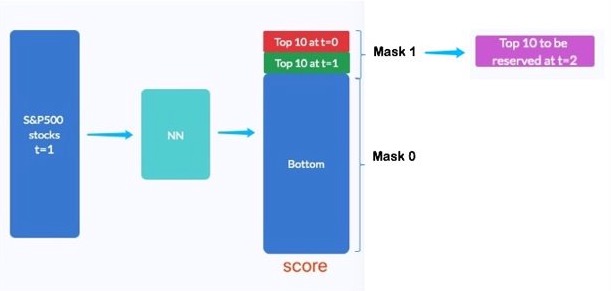}\\
  Figure 6 :Mask network for general stock selection with turnover control
\end{center}

\begin{center}
  \includegraphics[scale=0.8]{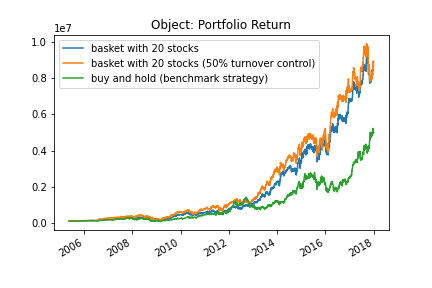}\\
  Figure 7 : Comparison of general selection model with and without turnover control
\end{center}

\subsection{Action Control on Stock Shares}
We can directly control the number of shares invested in each stocks. In this setting, decreasing the transaction costs is simply equivalent to decrease the change of stock shares. We change the activation function in output layer of actor-NN to be tanh(), which yields value with range from -1 to 1. Since our initial endowment is 100K US dollars, we can restrict the maximum number of stock shares per stock to be 100. Thus, the optimal stock shares generated equal the multiplication of 100 and the output of tanh() function. The new model setting can also help to achieve high performance result, as shown in the following result.
\begin{center}
  \includegraphics[scale=0.8]{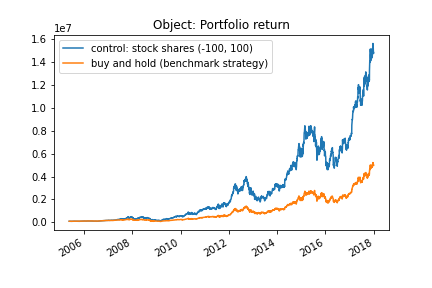}\\
  Figure 7 : Direct control on stock shares.
\end{center}

\section{Conclusions}
This paper introduces concepts and implementations of Deep Direct Reinforcement Learning framework (DDRL) for financial signal representation and portfolio trading. With risk \& reward models identified, we set up the objective of maximizing ultimate reward as absolute return in portfolio value. To build the DDRL framework, a recurrent neural net is introduced to handle sequential data representation, following by two two-layer deep neural network for decision learning. Method of Task Aware Back Propagation Through Time is introduced to address the issue of recurrent derivation and gradient vanishing. 

To compare performance of DRL against S\&P 500 index, the algorithm is tested by trading portfolios of selected stocks and comparing the result with the strategy of buying and holding an ETF named SPY, which tracks S\&P500 Index, at a time range between 2006 and 2019. The result appears to be promising with DRL agents outperforming market (return from Buy and Hold of SPY) return by an amount of \$250,000. 

However, to address the limitations related to trading cost and portfolio size, we introduce mask for general selection, set a threshold of the stocks turnover, make stock shares as action control instead of weights, and change objective to Sharpe ratio. To the best of our knowledge, this
is the first attempt to use the DRL for auto trading on a large pool of stocks. In the future, there could be room for further improvement on more sophisticated portfolio management.

%% file: main.bbl
\begin{thebibliography}{24}
\providecommand{\natexlab}[1]{#1}
\providecommand{\url}[1]{\texttt{#1}}
\expandafter\ifx\csname urlstyle\endcsname\relax
  \providecommand{\doi}[1]{doi: #1}\else
  \providecommand{\doi}{doi: \begingroup \urlstyle{rm}\Url}\fi

\bibitem[Deng et~al.(2017)Deng, Bao, Kong, Ren, and Dai]{article4}
Deng, Y., Bao, F., Kong, Y., Ren, Z., and Dai, Q.
\newblock “deep direct reinforcement learning for financial signal
  representation and trading”.
\newblock \emph{IEEE Trans. Neural Netw., vol. 28, no. 3, pp. 653 - 664}, 2017.

\bibitem[Du \& Jinjian~Z.(2009)Du and Jinjian~Z.]{article13}
Du, X. and Jinjian~Z., e.~a.
\newblock "algorithm trading using q-learning and recurrent reinforcement
  learning".
\newblock \emph{http://cs229.stanford.edu/proj2009/LvDuZhai.pdf}, 2009.

\bibitem[Fister et~al.(2019)Fister, Mun, Jagriˇc, and Jagriˇc]{article3}
Fister, D., Mun, J., Jagriˇc, V., and Jagriˇc, T.
\newblock “deep learning for stock market trading: A superior trading
  strategy”.
\newblock \emph{Neural Network World, 29(3):151–171}, 2019.

\bibitem[Grondman et~al.(2012)Grondman, Vaandrager, Busoniu, Babuska, and
  Schuitema]{article6}
Grondman, I., Vaandrager, M., Busoniu, L., Babuska, R., and Schuitema, E.
\newblock "efficient model learning methods for actor–critic control".
\newblock \emph{IEEE Transactions on Systems, Man, and Cybernetics, Part B
  (Cybernetics), vol. 42, no. 3, pp. 591-602}, 2012.

\bibitem[Han \& Liu(2016)Han and Liu]{han2016bootstrap}
Han, J. and Liu, Q.
\newblock Bootstrap model aggregation for distributed statistical learning.
\newblock In \emph{Advances in Neural Information Processing Systems}, pp.\
  1795--1803, 2016.

\bibitem[Han \& Liu(2017)Han and Liu]{han2017stein}
Han, J. and Liu, Q.
\newblock Stein variational adaptive importance sampling.
\newblock \emph{In Uncertainty in Artificial Intelligence}, 2017.

\bibitem[Han \& Liu(2018)Han and Liu]{han2018stein}
Han, J. and Liu, Q.
\newblock Stein variational gradient descent without gradient.
\newblock \emph{arXiv preprint arXiv:1806.02775}, 2018.

\bibitem[Han et~al.(2017)Han, Zhang, and Zhang]{han2017high}
Han, J., Zhang, H., and Zhang, Z.
\newblock High efficiently numerical simulation of the tdgl equation with
  reticular free energy in hydrogel.
\newblock \emph{arXiv preprint arXiv:1706.02906}, 2017.

\bibitem[Han et~al.(2018)Han, Lombardo, Schroers, and Mandt]{han2018deep}
Han, J., Lombardo, S., Schroers, C., and Mandt, S.
\newblock Deep probabilistic video compression.
\newblock \emph{arXiv preprint arXiv:1810.02845}, 2018.

\bibitem[Han et~al.(2020)Han, Ding, Liu, Torresani, Peng, and
  Liu]{han2020stein}
Han, J., Ding, F., Liu, X., Torresani, L., Peng, J., and Liu, Q.
\newblock Stein variational inference for discrete distributions.
\newblock \emph{arXiv preprint arXiv:2003.00605}, 2020.

\bibitem[Han et~al.(2021)Han, Min, Han, Li, and Zhang]{han2021disentangled}
Han, J., Min, M.~R., Han, L., Li, L.~E., and Zhang, X.
\newblock Disentangled recurrent wasserstein autoencoder.
\newblock \emph{arXiv preprint arXiv:2101.07496}, 2021.

\bibitem[Henrique et~al.(2019)Henrique, Sobreiro, and Kimura]{article1}
Henrique, B.~M., Sobreiro, V.~A., and Kimura, H.
\newblock “literature review: machine learning techniques applied to
  financial market prediction”.
\newblock \emph{Expert Syst. Appl. vol.124, pp.226–251}, 2019.

\bibitem[Jiang et~al.(2017)Jiang, Xu, and J.J.]{article9}
Jiang, Z., Xu, J., and J.J., L.
\newblock "a deep reinforcement learning framework for the financial portfolio
  management problem".
\newblock \emph{https://arxiv.org/abs/1706.10059}, 2017.

\bibitem[Jin \& El-Saawy(2016)Jin and El-Saawy]{article14}
Jin, O. and El-Saawy, H.
\newblock "portfolio management using reinforcement learning".
\newblock \emph{https://arxiv.org/pdf/1909.09571.pdf}, 2016.

\bibitem[Lee et~al.(2007)Lee, Park, O, Lee, and Hong]{article7}
Lee, J.~W., Park, J., O, J., Lee, J., and Hong, E.
\newblock "a multiagent approach to $q$-learning for daily stock trading".
\newblock \emph{IEEE Transactions on Systems, Man, and Cybernetics - Part A:
  Systems and Humans, vol. 37, no. 6, pp. 864-877}, 2007.

\bibitem[Liang et~al.(2018)Liang, Chen, Zhu, Jiang, and Li]{article10}
Liang, Z., Chen, H., Zhu, J., Jiang, K., and Li, Y.
\newblock "adversarial deep reinforcement learning in portfolio management".
\newblock \emph{https://arxiv.org/abs/1808.09940}, 2018.

\bibitem[Lillicrap \& Hunt(2016)Lillicrap and Hunt]{article12}
Lillicrap, T. and Hunt, J.J., e.~a.
\newblock "continuous control with deep reinforcement learning".
\newblock \emph{https://arxiv.org/abs/1509.02971}, 2016.

\bibitem[Lombardo et~al.(2019)Lombardo, Han, Schroers, and
  Mandt]{lombardo2019deep}
Lombardo, S., Han, J., Schroers, C., and Mandt, S.
\newblock Deep generative video compression.
\newblock In \emph{Advances in Neural Information Processing Systems}, pp.\
  9283--9294, 2019.

\bibitem[Moody \& Saffell(1999)Moody and Saffell]{article2}
Moody, J. and Saffell, M.
\newblock “reinforcement learning for trading,”.
\newblock \emph{Advances in Neural Information Processing Systems, vol. 11, pp.
  917-923, MIT Press}, 1999.

\bibitem[Moody \& Saffell(2001)Moody and Saffell]{article8}
Moody, J. and Saffell, M.
\newblock "learning to trade via direct reinforcement".
\newblock \emph{in IEEE Transactions on Neural Networks, vol. 12, no. 4, pp.
  875-889}, 2001.

\bibitem[Schulman et~al.(2017)Schulman, Wolski, Dhariwal, Radford, and
  Klimov]{article11}
Schulman, J., Wolski, F., Dhariwal, P., Radford, A., and Klimov, O.
\newblock "proximal policy optimization algorithms".
\newblock \emph{https://arxiv.org/abs/1707.06347}, 2017.

\bibitem[Sutton \& Barto(1998)Sutton and Barto]{article5}
Sutton, R.~S. and Barto, A.~G.
\newblock “reinforcement learning: An introduction”.
\newblock \emph{Cambridge, MA, USA: MIT Press}, 1998.

\bibitem[Weijs(2018)]{article15}
Weijs, L.
\newblock "reinforcement learning in portfolio management and its
  interpretation".
\newblock \emph{Erasmus Universiteit Rotterdam}, 2018.

\bibitem[Yang \& Liu(2020)Yang and Liu]{article16}
Yang, H. and Liu, X.Y., e.~a.
\newblock "deep reinforcement learning for automated stock trading: An ensemble
  strategy".
\newblock \emph{ACM International Conference on AI in Finance.}, 2020.

\end{thebibliography}
